# Analysis of bistability at the coupling between waveguide and whispering gallery modes of a nonlinear hemicylinder

**Henrik Parsamyan[1,2], Khachik Sahakyan[1] and Khachatur Nerkararyan[1]**

[1] Department of Microwave Physics and Telecommunication, Yerevan State University, 1 Alex Manoogian str., Yerevan 0025, Armenia

[2] Center for Nanoscience and Technology, Institute of Chemical Physics, NAS RA, 5/2 P. Sevak str., Yerevan 0014, Armenia

Email: hparsamyan@ysu.am

**Abstract**

The optical bistability caused by the coupling between modes of the parallel-plate waveguide and a nonlinear hemicylindrical crystal is studied using theoretical and numerical analysis. In such a system a waveguide channel is parallelly coupled to whispering gallery modes of a hemicylindrical microresonator ensuring bistable behaviour at input intensities of the order of a few MW/cm$^2$. The characteristic minimum switching time of the system (around 30 ps) can be controlled by varying the thickness of the metal layer which couples the waveguide and whispering gallery modes. This is conditioned by the change of the quality $Q$-factor, as well as the coupling coefficient of the resonator. The main advantages of the system are fabrication simplicity, small sizes of the order of 3 μm and the possibility of adjusting the processes by making use of the electro-optical effect.

## Introduction

Realization of all-optical devices such as switches, logic gates and optical memory elements in micrometre scales can open up new prospects for next-generation optical communication and information processing with increased speed and bandwidth[1,2]. The last decade saw unprecedented growth of data generated and driven by Artificial Intelligence, the Internet of Things, Biotechnology and more. In fact, this growth was by magnitude faster than Moor's law predicting the processing growth which was required to analyze and store this data. To deal with such a huge amount of data, well-organized multicore processors and their interconnects between cores and other chips for parallel processing have been introduced. However, the main bottleneck in the server architecture is no longer the CPU performance but the interconnect within the chip, between the chips of the system and the memory access architecture[3,4]. There is already successful integration of photonics chip with electronics with strong evidence that optical memories and most importantly optical interconnection will provide much better performance per Watt, much lower latency and higher throughput [2,5,6]. Similar problems exist in networking: recent 400G and 800G network standards require doing as many tasks on the optical environment as possible before converting signals to electrical interface. There are already optics-based switch interfaces and the industry is trying and offload network processing to optics[2,7,8]. This requires optical ternary content addressable memories[9] that can return values in a single clock edge at several GHz clock rates which leads to well below-one-nanosecond latency requirement. Successful realization of bit addressable, SRAM-like, ultrafast optical memories and optical logic cells, where light is utilized to control the processes, requires making use of nonlinear properties of materials. The operation principle of all-optical devices aimed to address the above-mentioned issues (switches, transistors, logic gates, and memory elements) is based on the phenomenon of optical bistability[1], where the intensity-dependent nonlinear refractive index $n(I)$ of materials is exploited to allow for switching between two stable states by using light itself also known as the Kerr effect[10]. Metamaterials [11,12], plasmonic nanoantennas [13] and whispering gallery mode (WGM) microresonators[10,14–16] have been considered for enhanced optical bistability as they can increase the incident light by several orders of magnitude thus enabling operations at relatively low input powers. Particularly, waveguide-integrated microring resonators, being one of the main building blocks of integrated photonic chips and characterized by high and ultrahigh quality $Q$-factors, were investigated for generation of frequency combs and higher harmonics[17,18], electro-optical[19,20], as well as all-optical switching exploiting the optical bistability phenomenon, where a Kerr nonlinear material is used as a resonator host medium[15,21,22]. Although silicon is the main material for integrated photonics building blocks, III-V-group semiconductors such as GaAs and InP also are promising candidates for investigation of optical bistability due to their pronounced nonlinear optical properties. Bistable response of an optical system can be commonly analyzed by graphical-based post processing method. However, the main drawback of this method is that spatial variations of the local intensity inside the nonlinear medium and hence refractive index changes are neglected[12]. As a result, the modulation of the refractive index is associated with the average or maximum intensity inside the medium and homogeneous $n(I)$ is assumed. A more accurate approach is using numerical modeling tools based on finite-difference time-domain (FDTD) or finite element method (FEM) analysis where the spatially inhomogeneous distribution of $n(I)$ of the nonlinear system is taken into account.

In this article, we investigate the cavity-enhanced optical bistability of a system composed of a hemicylindrical WGM resonator coupled to a metal-insulator-metal (MIM) planar waveguide. The resonator host medium is GaAs with noticeable nonlinear properties. We bypass the conventional graphical method-based approach for analyzing the optical bistability by conducting numerical simulations based on the FEM. Such an approach allows one to take into account spatial variations of the resonator intensity, so that the refractive index of the resonator medium is a function of the *local* intensity inside the resonator. A hysteresis-like behaviour of the system is observed when the input intensity is either increased and decreased resulting in two distinct values of the waveguide transmittance. The analysis shows that the switching threshold values of the input intensity of ON-OFF and OFF-ON states are about 6.2 and 3.8 MW/cm$^2$, respectively. The main advantage of the proposed bistable device is that not only the resonator integrated with an optical planar waveguide is used, but also it can be integrated with other optical components. Moreover, the system performance is fully controlled by the waveguide input signal and the bistable behaviour can be detected by measuring the signal at the waveguide output. In addition, the working wavelength of the system can be tuned by using electro-optical effects.

**Results and Discussion**

The system is composed of a hemicylindrical WGM microresonator coupled with a MIM waveguide, schematically illustrated in Fig. 1(a). The radius of the hemicylinder is 3 μm, the thickness of the metal layer between the resonator and waveguide is 35 nm and the width of the waveguide is 500 nm. The thickness of the bottom metal layer of the waveguide is set at 500 nm to prevent radiation losses to the substrate of the system. As the waveguide and hemicylinder WGMs couple via the top metal layer, the thickness of the bottom metal does not influence the system performance. GaAs is used as a resonator host medium and Ag-SiO$_2$ were employed as metal-insulator for the waveguide. The intensity-dependent refractive index of GaAs is defined as

$$n_{\text{GaAs}} = n_0 + n_2 I_{\text{res}}, \qquad (1)$$

where $n_0$ is the linear or low-intensity refractive index, $n_2$ is the nonlinear refractive index coefficient and $I_{\text{res}}$ is the intensity inside the resonator. Throughout all analysis, constant values of $n_0 = 3.47$, $n_2 = -3\times10^{-13}$ cm$^2$/W and $n_{\text{SiO2}} = 1.5$ were used; which are typical for GaAs and SiO$_2$ in the near-infrared around 1 μm [23,24].The wavelength-dependent refractive index of silver is taken from [25]. The surrounding medium of the resonator is air. Since in the studied spectrum $n_2 < 0$, for convenience equation (1) is rewritten as follows:

$$n_{\text{GaAs}} = n_0 - |n_2| I_{\text{res}} = n_0 - \Delta n, \qquad (2)$$

where $\Delta n$ is the refractive index change due to Kerr nonlinearity.

Whispering gallery modes of the system are generally characterized by radial and azimuthal mode numbers[20]. The transmission spectrum of the TE(1,57) fundamental whispering gallery mode of the resonator with the resonant wavelength of $\lambda_0 = 1037.84$ nm and the distribution of the $E_z$ component of the electric field are shown in Fig. 1 (b) and (c), respectively. The full width at half maximum (FWHM) $\delta\lambda \sim 0.104$ nm, corresponding to the $Q$-factor $\sim 9980$, which is competitive with conventional Fabry-Pérot (FP) resonators.

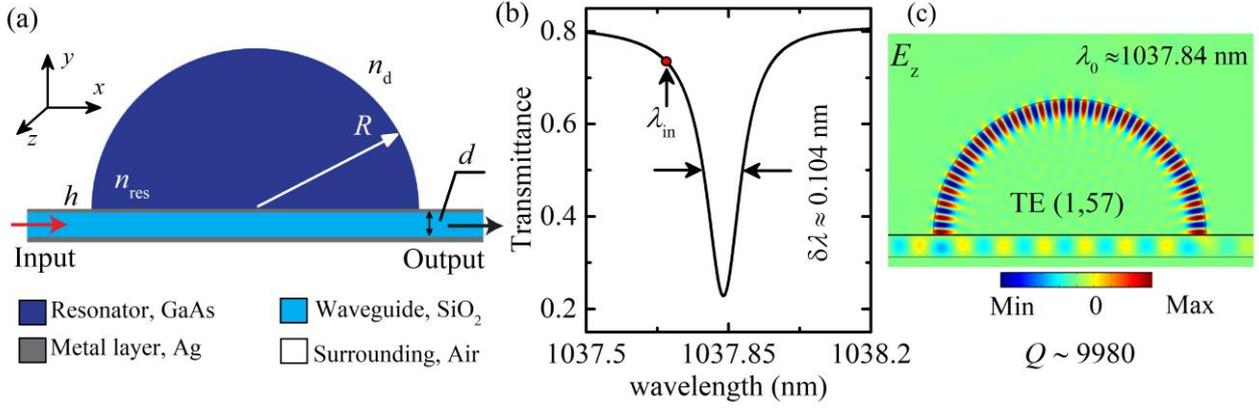

**Figure 1.** (a) Schematic of the hemicylindrical resonator coupled to the MIM waveguide. (b) Transmittance spectrum and (c) distribution of the electric field $E_z$ component of TE (1,57) mode with $Q$-factor ~ 9980.

One of the key factors to observe the optical bistability is defining the incoming wavelength $\lambda_{in}$ which should be shifted from the original resonant wavelength and must satisfy to the following condition [26]:

$$|\lambda_0 - \lambda_{in}| > \sqrt{3}\delta\lambda/2, \quad (3)$$

where $\lambda_0$ is the linear resonant wavelength and $\delta\lambda$ is the FWHM of the resonant curve. Condition (3) defining the input wavelength at which two output steady-states are observed has a common nature and is characteristic for resonant nonlinear oscillations in general[26,27]. The relevance of this condition has been proven in many studies of nonlinear systems, including individual nanoantennas, metasurfaces, plasmonic and WGM resonators[12,28,29]. The incoming wavelength in the studied case must be chosen according to the expression $\lambda_{in} < \lambda_0 - \sqrt{3}\delta\lambda/2$ due to the negative nonlinearity $n_2 < 0$. Taking into account the latter and the fact that in the low-intensity regime the resonant wavelength $\lambda_0 = 1037.84$ nm and $\delta\lambda = 0.104$ nm, one can expect to ensure optical bistability at incoming wavelengths below 1037.75 nm.

To identify the approximate range of input intensities assuring optical bistability, we set the incoming wavelength at 1037.7 nm and plotted the dependence of the *maximum* intensity enhancement $\eta = I_{r,max}/I_{in}$ inside the resonator versus the change of the resonator refractive index $\Delta n$ in the range from $10^{-5}$ to $10^{-3}$. Note that such value of the incident wavelength is chosen to ensure low switching thresholds and a moderate range of bistability. The results are shown in Fig. 2(a) left axis. Fig. 2 (a) right axis shows the transmittance $T$ of the system as a function of $\Delta n$. These are two main functions used to retrieve the bistability hysteresis curve based on the graphical method. Once the functional dependencies of $\eta_{max}$ and $T$ on $\Delta n$ were obtained via FEM simulations, $T(I_{in})$ curve can be derived as follows: the input intensity is related to the resonator intensity and enhancement factor as $I_{in} = I_{r,max}/\eta_{max}$; and the maximum intensity $I_{r,max}$ inside the resonator required for the introduced refractive index change can be calculated according to the relation $I_{r,max} = \Delta n/n_2$ [see eq. (2)]. Thus, the input intensity and the refractive index change are related as $I_{in} = \Delta n/(n_2\eta_{max})$. Finally, juxtaposing $I_{in}(\Delta n)$ and $T(\Delta n)$ one obtains the bistability curve of the resonator-waveguide system[30]. One sees that the resonance at the shifted input wavelength occurs for $\Delta n \sim 4.8 \cdot 10^{-4}$. By choosing a range of $\Delta n$ which includes this resonant value and taking into account the corresponding intensity enhancement factors, the input intensity range for observing optical bistability can be estimated according to expression $I_{in} = \Delta n/(n_2\eta_{max})$. Thus, considering $\Delta n$ interval ~ $(2 \cdot 10^{-5}; 6 \cdot 10^{-4})$ with corresponding intensity enhancement factors $\eta_{max}$ ~123 and 495 at the range boundaries, one estimates the range of input intensities of the optical bistability to be approximately 0.5-4 MW/cm$^2$. Fig. 2(b) is a confirmation of this representing the transmission spectra of the system in the linear case (black curve) and at input intensities 1, 2 and 3 MW/cm$^2$ (red, blue and purple curves, respectively) derived via nonlinear numerical simulations. One sees that in the linear regime (i.e., the input intensity is very low) the line shape of the resonance is strictly Lorentzian, whereas higher input intensities deform the line shape into triangular with a blueshift of the peak wavelength. At shorter wavelengths, the intensity inside the resonator is not sufficient to modulate its refractive index. Increasing the input wavelength and reaching some point, the resonator intensity becomes high enough to decrease the refractive index (due to a negative $n_2$) and hence matching between the resonant wavelength of the cavity WGM and the incoming wavelength is ensured. Further increase of the wavelength gradually detunes the resonance. Note that both the linear refractive index and Kerr nonlinear coefficient are considered to be wavelength-independent in the narrow wavelength range.

The hysteresis loop of the transmittance at the waveguide end as a function of the input intensity derived through the so-called graphical method based on the functional dependence of $\eta$ and $T$ on $\Delta n$ presented in Fig. 2(a) and eq. (2) is shown in Fig. 3(a). The upper stable arm of the hysteresis can be viewed as the ON state, whereas the lower one - as the OFF state.

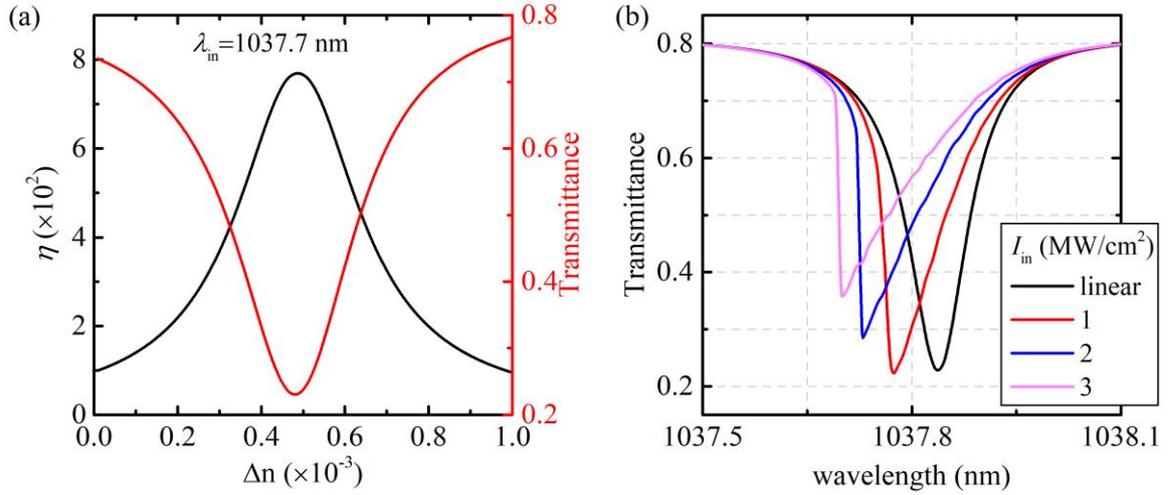

**Figure** 2. (a) Maximum intensity enhancement $\eta$ inside the resonator (left axis) and the transmittance (right axis) as a function of the GaAs refractive index change $\Delta n = n_2 I_{r,max}$ at the input wavelength $\lambda_{in} = 1037.7$ nm. (b) Transmission spectrum of the system in the linear regime (black) and for input intensities of 1 (red), 2 (blue) and 3 (purple) in MW/cm$^2$.

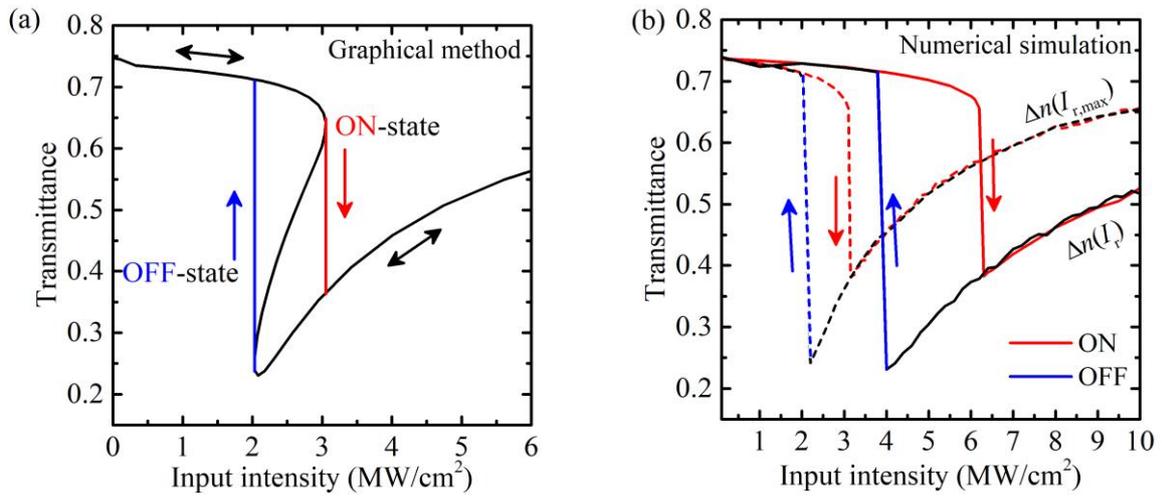

**Figure 3**. Optical bistability of the hemicylindrical resonator-MIM waveguide system. (a) Transmittance as a function of the input intensity obtained via a common graphical method using results in Fig. 2(a). (b) Optical bistability curves obtained through full-wave numerical FEM simulations by using intensity-dependent refractive index defined by the maximum $I_{r,max}$ and local $I_r$ intensity inside the resonator. The input wavelength is $\lambda_{in} = 1037.7$ nm.

Although the graphical method-based derivation of the hysteresis of the optical bistability is widely used due to the difficulties of the nonlinear numerical computations, it assumes that the refractive index of the nonlinear medium varies uniformly depending on the intensity and spatial variations of the refractive index are not taken into account. However, in general, the distribution of the wave energy, and hence the spatial variation of the refractive index is not uniform. To deal with this issue, FEM-based numerical simulations were conducted, where the intensity-dependent refractive index is a function of the local intensity inside the resonator. First, in order to make sure that our nonlinear simulations are correct and to compare the results with those of the graphical method-based analysis, we defined a refractive index change function linked to the maximum resonator intensity $\Delta n(I_{r,max})$ and starting from $I_{in} = 0.1$ MW/cm$^2$, gradually increased the input intensity. This resulted in the first upper (ON) branch of the hysteresis loop. The lower (OFF) branch is obtained by setting the input intensity to $I_{in} = 10$ MW/cm$^2$ and gradually decreasing it. Results are shown in Fig. 3(b) by the red and black dashed lines, respectively. One sees that the optical bistability loops obtained via the graphical method and nonlinear simulations correspond to each other. Note that in the nonlinear numerical calculations converged solutions from the previous step are used for each following step.

Although analysis based on the $\Delta n(I_{r,max})$ gives one an idea about the bistable behaviour of the system, it will obviously over-estimate the range of input intensities. Red and black solid lines in Fig. 3 (b) obtained via the same procedure as above represent the hysteresis loop of the transmittance of the system as a function of the input intensity, where the intensity-dependent refractive index change is defined by the *local* intensity $\Delta n(I_r)$ inside the resonator. The transmittance for the upper branch maintains approximately constant as the input intensity increases and approaching the first threshold intensity $I_{ON\text{-}OFF} \sim 6.2$ MW/cm$^2$, instantly drops down to the OFF state. On the contrary for the lower hysteresis branch, as the input intensity decreases reaching the second threshold intensity $I_{OFF\text{-}ON} \sim 3.8$ MW/cm$^2$ corresponding to the critical coupling of TE mode of the waveguide and WGM of the resonator, the transmittance jumps from the OFF to the ON state. The ON/OFF and OFF/ON ratios of the system are calculated to be 1.8 and 3.1, respectively.

The minimum switching time of the system is defined as $\tau_c = Q\lambda_0/c$, where $Q$ is the quality factor of the resonator mode, $\lambda_0$ is the resonant wavelength and $c$ is the speed of light[12]. Taking into account the results in Fig. 1, one estimates the minimum switching time of the system to be around 34 ps. The metal thickness $h = 35$ nm in the above analysis was chosen to ensure the highest field enhancement inside the resonator. However, the $Q$-factor of the resonator can be changed by varying the metal thickness, as shown in Fig 4(a). Here, systems with 25, 35 and 50 nm-thick metal layers are considered, with corresponding $Q$-factors of ~ 5133, 9984 and 17480. Thus, the minimum switching time of the system can be adjusted by varying the thickness of the metal. For instance, $\tau_c$ for $h = 25$ nm is half $\tau_c$ at $h = 35$ nm. Optical bistability hysteresis loops of the system composed of 25 and 50-nanometer-thick Ag layers (All else is as in Fig. 1) are compared in Fig. 4 (b). Although $\tau_c$ for $h = 25$ nm is relatively short, the field enhancement inside the resonator is quiet low and consequently the range of input intensities to observe optical bistability is higher. It is seen that in both cases the overall system performance is worse than in the previous case, taking into account the ON / OFF and OFF / ON ratios and the range of bistability intensity. One sees that the resonant wavelength of the TE(1,57) mode is slightly changed as the metal thickness between the waveguide and the hemicylinder varies. To reveal the physical origin of such a shift, one should examine peculiarities of the formation of WGMs in the suggested system. Unlike conventional WGM resonators, where the resonator modes are formed by total internal reflections from the resonator surface, here the WGMs are formed by both continuous total internal reflections from the curved boundary of the hemicylinder and reflections from the metal layer on the bottom of the hemicylinder. Thus, the system combines the features of WGM and FP resonators, where two mirrors lying in the same plane are connected by a curved path. Since the Ag thickness varies in the nanometre range (up to 50 nm), the reflection phase is also changed causing a shift in the resonant wavelength [31].

From the point of view of practical implementation, the bistable behaviour of the system should also be observed when the hemicylinder is surrounded by a fixing dielectric instead of air. Such a fixture will prevent the hemicylinder from mechanical damage and will essentially increase the stability of the system to thermal influence. Solid lines in Fig. 5 depict the hysteresis loop of the transmittance versus input intensity for a system composed of a hemicylinder surrounded by a lossless dielectric with refractive index $n_d = 1.48$. Here again, the fundamental TE(1,57) mode is considered whose linear resonance is shifted at ~ 1039 nm due to the presence of a dielectric layer outside of the hemicylinder and the wavelength of the driving field is 1038.89 nm. Ratios of transmission coefficients of ON/OFF and OFF/ON states are slightly decreased due to the presence of the dielectric fixture which decreases of the potential barrier at the hemicylinder-surrounding interface thus increasing radiation losses. One of the main applications of an optical nonlinear system is switching. The operation principle of the system as an optical switch is shown in Fig. 5 by dashed lines. The exterior of hemicylinder is a dielectric with $n_d = 1.48$. One sees that by choosing the wavelength of the input driving signal, it is possible to obtain hysteresis or step-like behavior[26].

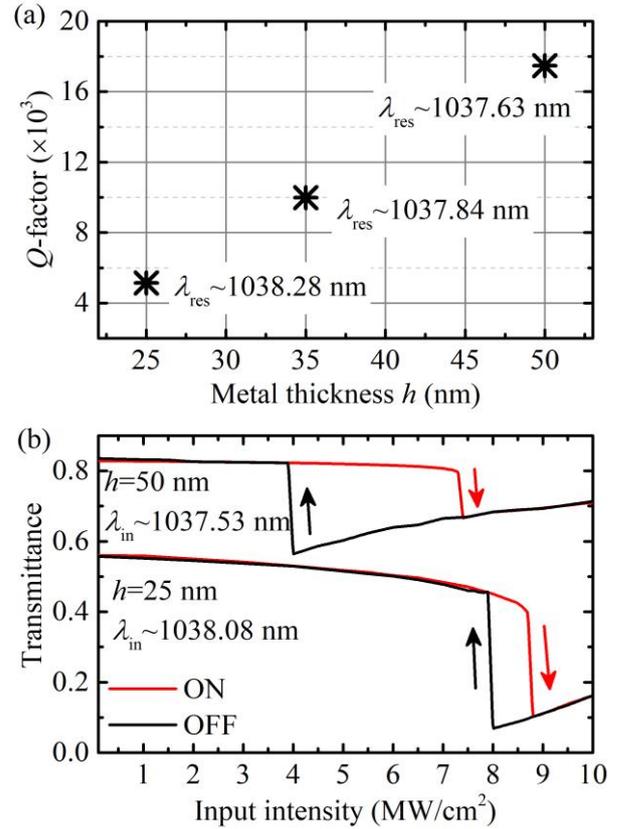

**Figure 4**. (a) Dependence of the $Q$-factor of the resonator on the silver thickness (25, 35 and 50 nm) with the corresponding resonant wavelengths. (b) Transmittance of the system as a function of the input intensity for silver thickness - input wavelength of 25 nm - 1038.08 nm and 50 nm - 1037.53 nm.

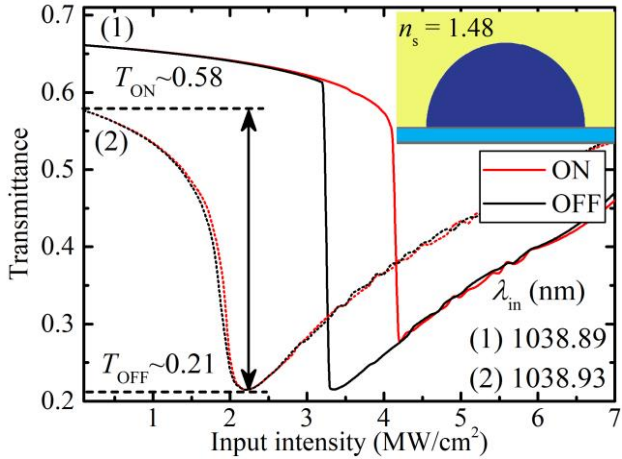

**Figure 5**. The transmittance of the system as a function of the input intensity in the case of a hemicylinder surrounded by a dielectric with $n_d = 1.48$ at driving wavelengths of 1038.89 nm (solid lines) and 1038.93 nm (dashed lines). The calculation was performed for TE(1,57) mode. Red and black lines stand for ON and OFF branches, respectively.

Particularly, when the wavelength of an input signal is close to the limit defined by eq. (3), step-like behaviour of the transmittance as a function of the input intensity is observed. At low input intensities $T \sim 0.58$, whereas at $I_{in} \sim 2.2$ MW/cm$^2$ $T \sim 0.21$. Such functionality also follows from Fig. 2 (b). At the input wavelength of 1037.84 nm, which corresponds to the small-signal linear resonance, $T \sim 0.22$, while at the input intensity of 3 MW/cm$^2$ $T \sim 0.63$ (nonlinear regime).

Importantly, III-V group materials are also characterized by large electro-optical coefficients. Since the incoming wavelength for achieving optical bistability (nonlinear regime) is shifted from the resonant wavelength at the linear regime, the bistability wavelength can be adjusted by using the electro-optical effect. Finally, although in conventional integrated photonics Si is widely used, the Kerr nonlinear coefficient of this material is more than ten times smaller compared with that of GaAs[10]. This means that using Si as a host medium for the resonator whose sizes are of the order of the operating wavelength will require higher input intensities to observe optical bistability. On the other hand, higher intensity enhancement factors and thus larger $Q$-factors (or resonator sizes) will be needed for operation at relatively low input intensities. In contrast to conventional high-$Q$-factor WGM resonators, the suggested system is much less sensitive to environmental conditions such as the surrounding refractive index or temperature changes due to $Q$-factors of the order of $10^4$. In addition, adding a dielectric fixture will further improve the stability of the system by increasing heat dissipation.

Ideas presented here for the two-dimensional model of the device can be further applied to a more practical three-dimensional system. However, one should bear in mind that the system size along the hemicylinder axis should be much larger than the wavelength in the resonator. Additionally, radiation losses from the hemicylinder surface will be increased.

The fabrication process of the proposed design of an optical bistable device can be implemented in two main steps. The fabrication of the waveguide part is rather straightforward and includes applying metal and dielectric layers to the semiconductor layer. After the thickness of the semiconductor layer is reduced to the resonator sizes, hemicylinder can be created by a well-introduced technique of the fabrication of semiconductor microlenses or anisotropic chemical etching [32,33].

## Conclusions

Thus, in a coupled system composed of a planar waveguide and hemicylindrical nonlinear crystal favourable conditions for observing optical bistability arise. Unlike conventional microresonators, whispering gallery modes of the hemicylindrical resonator here couple with a planar waveguide parallelly in two regions, which extends the possibilities of controlling the optical processes. The optical bistability in the hemicylindrical microresonator of 3 μm-radius is achieved at input intensities of the order of a few MW/cm$^2$. The $Q$-factor of the resonator and thus the characteristic switching time of the system can be controlled by changing the nanometer-thin metal film coupling the planar waveguide and hemicylinder whispering gallery modes. At the studied conditions the minimum switching time of the system is around 30 ps. The fabricated process of the device is similar to the well-introduced technique of semiconductor microlenses. The simplicity of fabrication, small sizes and the possibility of adjusting the optical processes by electro-optical methods, as well as integrability with other optical components are favourable factors for using such a system.

## Acknowledgments

This work was supported by a scientific research grants through the SC of MESCS of Armenia (21AG-1C061 and 21DP-2J017).